\documentclass[iop]{emulateapj}
\usepackage{apjfonts}

\slugcomment{Accepted for publication in ApJ Letters}
\shortauthors{Ree et~al.}
\shorttitle{UV color--color relation of early-type galaxies}

\begin{document}

\title{Ultraviolet color--color relation of early-type galaxies at $0.05<\lowercase{z}<0.12$}

\author{
Chang~H.~Ree\altaffilmark{1},
Hyunjin~Jeong\altaffilmark{1,2},
Kyuseok~Oh\altaffilmark{2},
Chul~Chung\altaffilmark{2},
Joon~Hyeop~Lee\altaffilmark{1},
Sang~Chul~Kim\altaffilmark{1},
Jaemann~Kyeong\altaffilmark{1}
}

\altaffiltext{1}{Korea Astronomy and Space Science Institute, Daejeon 305-348, Republic of Korea (e-mail : chr@kasi.re.kr)}
\altaffiltext{2}{Department of Astronomy, Yonsei University, Seoul 120-749, Republic of Korea}

\begin{abstract}
We present the ultraviolet (UV) color--color relation of early-type galaxies (ETGs) in the nearby universe ($0.05<z<0.12$) to investigate the properties of hot stellar populations responsible for the UV excess (UVX). The initial sample of ETGs is selected by the spectroscopic redshift and the morphology parameter from the Sloan Digital Sky Survey (SDSS) DR7, and then cross-matched with the Galaxy Evolution Explorer (GALEX) Far-UV (FUV) and Near-UV (NUV) GR6 data. The cross-matched ETG sample is further classified by their emission line characteristics in the optical spectra into quiescent, star-forming, and AGN categories. Contaminations from early-type spiral galaxies, mergers, and morphologically disturbed galaxies are removed by visual inspection. By drawing the $FUV-NUV$ (as a measure of UV spectral shape) $vs.$ $FUV-r$ (as a measure of UVX strength) diagram for the final sample of $\sim$3700 quiescent ETGs, we find that the ``old and dead'' ETGs consist of a well-defined sequence in UV colors, the ``UV red sequence'', so that the stronger UVX galaxies should have a harder UV spectral shape systematically. However, the observed UV spectral slope is too steep to be reproduced by the canonical stellar population models in which the UV flux is mainly controlled by age or metallicity parameters. Moreover, 2~mag of color spreads both in $FUV-NUV$ and $FUV-r$ appear to be ubiquitous among any subsets in distance or luminosity. This implies that the UVX in ETGs could be driven by yet another parameter which might be even more influential than age or metallicity.
\end{abstract}


\section{INTRODUCTION}

The ultraviolet excess (UVX) refers to the phenomenon that the early-type galaxies (ETGs) exhibit a rising flux with a decreasing wavelength from 2500{\AA} to the Lyman limit. Theoretical and observational evidence shows that the phenomenon originates from the hot helium-burning horizontal-branch (HB) stars and their progeny \citep{oco99,bro00}. One of the most important observational constraints to understanding this phenomenon was given by \citet{bur88}. They first explored the systematics of the UV radiation from the 31 local ETGs and the bulge of M31, and found that the UVX strength correlates with the nuclear spectral line index Mg$_2$ (and, weakly, with the central velocity dispersion and luminosity) in $quiescent$ galaxies. With the advent of the Galaxy Evolution Explorer (GALEX) satellite \citep{mart05}, a large area, deep UV imaging survey gathers high-quality data for an unprecedentedly large sample of ETGs in the nearby universe. With the GALEX UV data cross-matched with the Sloan Digital Sky Survey (SDSS), \citet{ric05} investigated the systematics of the ETGs at $z<0.2$, but found no correlation of the UVX strength against the Mg$_2$, D4000 indices or stellar velocity dispersion. On the other hand, \citet{don07} found a tight correlation between the GALEX UV color and the Mg$_2$ index among the morphologically sorted ETGs in the RC3 catalog \citep{dev91}. Recently, \citet{bur11} recovered the correlation of Burstein et~al. from the GALEX UV imagings combined with the high-quality data of the SAURON sample, and also with the SDSS data for a more distant ETG sample. They argued that such correlations could be seen among the $quiescent$ ETGs only when the contaminations of star-forming galaxies and low-quality data were much more stringently removed than was done in \citet{ric05}.

The UV color--color diagram is another powerful tool to investigate the properties of hot stellar populations responsible for the UVX. Particularly, the capability of the GALEX to observe the far-UV (FUV; 1528{\AA}) and near-UV (NUV; 2271{\AA}) bandpasses simultaneously \citep{mor05} let us securely estimate the UV spectral shape without instrumental biases. As demonstrated by \citet{dor95}, the UV color--color diagram, $FUV-NUV$ (as a measure of UV spectral shape) $vs.$ $FUV-r$ (or $FUV-V$; as a measure of UVX strength) can trace the UV spectral shape and amplitude. It is also useful to segregate different stellar population systems. \citet{ric05} employed this tool but found no systematic variation among their ETGs at $z<0.2$. \citet{ree07} presented another example to show that the brightest cluster galaxies (BCGs) are homogenous populations following an old stellar sequence with considerable amounts of color spread.

In this $Letter$, we expand significantly the sample size of previous studies from the latest data release of the GALEX and SDSS to construct the UV color--color relation of the ETGs in the nearby universe, and reexamine our current understanding of UVX.

\section{GALEX -- SDSS EARLY-TYPE GALAXY SAMPLE}

\begin{figure*}
\begin{center}
\epsscale{1.0}
{\plottwo{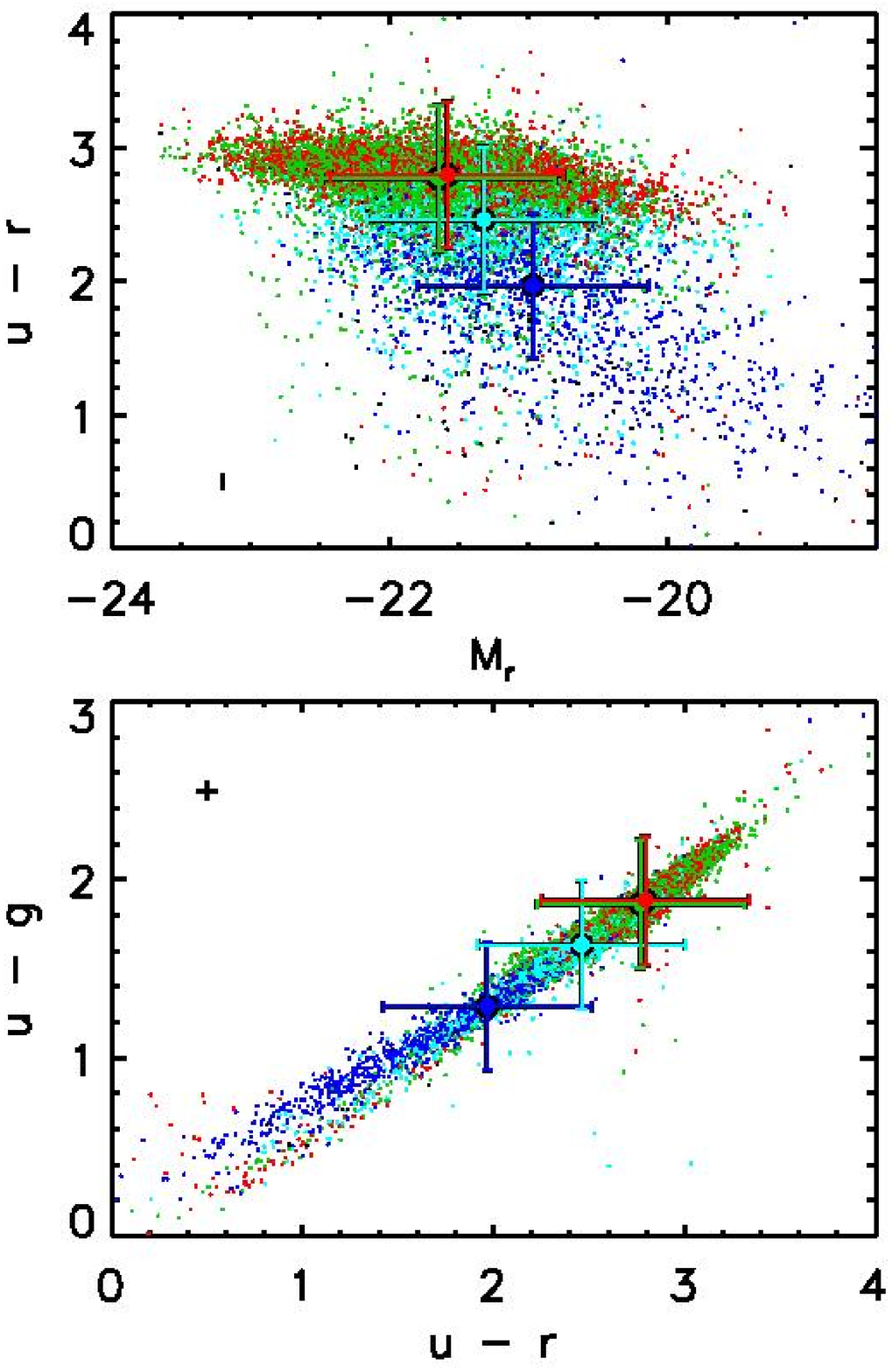}{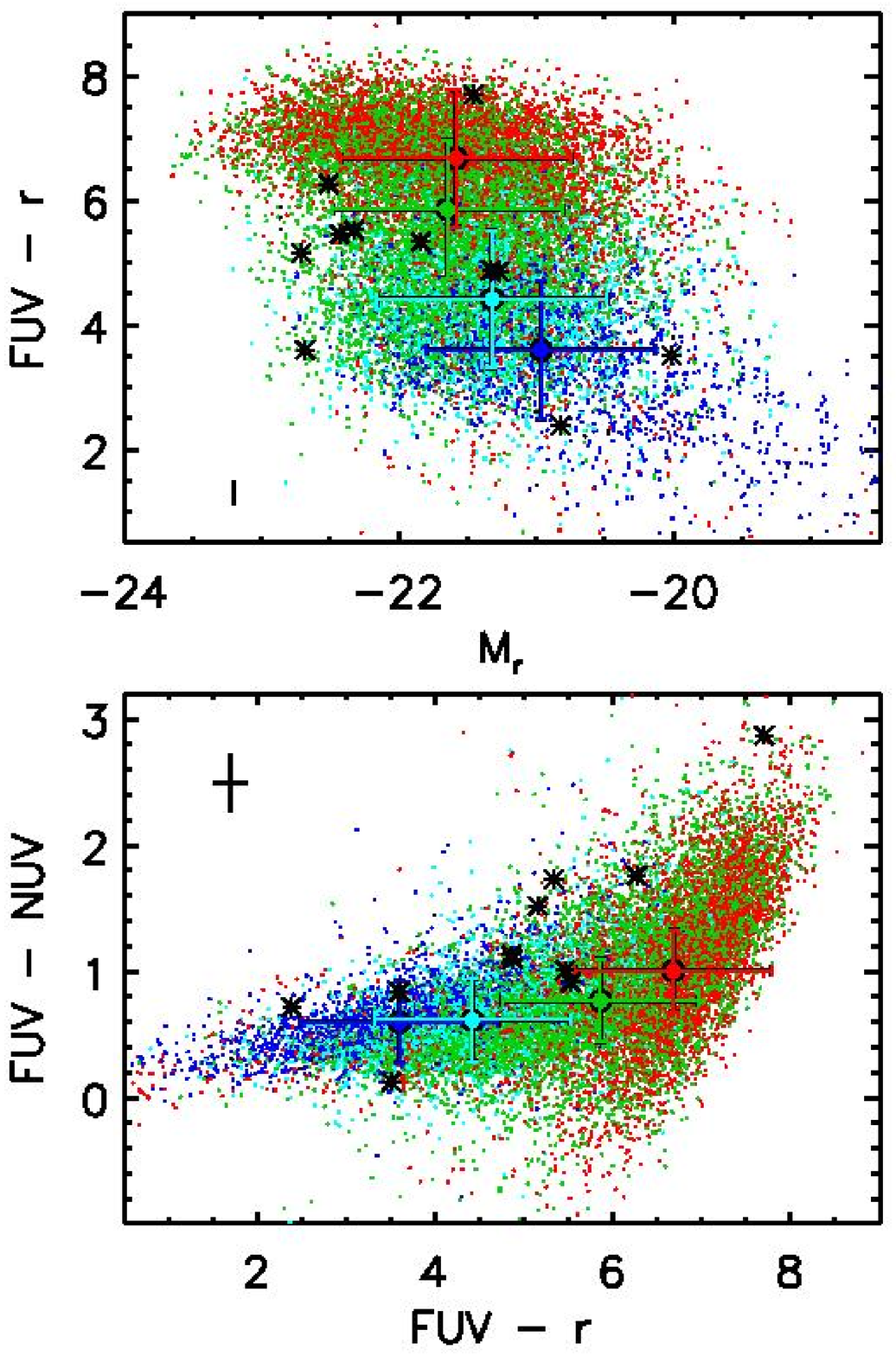}}
\end{center}
\caption{$(Left) :$ $u-r$ $vs.$ M$_r$ color--magnitude diagram ($top$) and $u-g$ $vs.$ $u-r$ color--color diagram ($bottom$) of the initial SDSS ETG catalog. $(Right) :$ $FUV-r$ $vs.$ M$_r$ color--magnitude diagram ($top$) and $FUV-NUV$ $vs.$ $FUV-r$ color--color diagram ($bottom$) of the GALEX--SDSS ETG cross-matched catalog. The symbols are color-coded according to the spectroscopic types (red: quiescent, green: AGN, cyan: composite, blue: star-forming) by the emission line diagnostics \citep{oh11}. Filled circles with error bars indicate the median colors and magnitudes with the standard deviation of each classified subgroup. Some E+A galaxies are found in our data and their UV colors are indicated by $asterisks$. The median observational errors are denoted by the black crosshairs. \label{fig1}}
\end{figure*}

The initial sample of ETGs is selected from the SDSS DR7 spectroscopic and photometric database in a rather simplistic way, with the following two criteria:
\begin{displaymath}
\begin{array}{c}
\mathrm{spectroscopic~redshift}: 0.05 < z < 0.12\\
\mathrm{morphology~parameter}: fracDeV_r > 0.99
\end{array}
\end{displaymath}
The search volume is limited as above in order to secure an unbiased sample from the SDSS spectroscopic survey. At $z<0.05$, the SDSS spectroscopy is incomplete for bright galaxies. The SDSS target selection algorithm of the main galaxy sample allocates the fibers to the galaxies $r=17.77$ or brighter \citep{sto02,str02}. This magnitude limit corresponds to $M_r \approx -21$ at $z=0.12$. Then, we measure the early-type morphology with the $fracDeV$ parameter in r-band. The $fracDeV$ is a fractional parameter that measures the amount of flux contribution by fitting a galaxy image to a de Vaucouleurs surface brightness profile \citep{dev48}. By selecting $fracDeV_r \approx 1$ galaxies, our sample secures purely spheroidal galaxies even though a few percent of the sample turn out to be disk galaxies with prominent bulges. Those contaminants are removed later from the final sample through visual inspection. The SDSS DR7 database\footnote{http://cas.sdss.org/dr7} returns an optical catalog of 94,796 galaxies with our querying parameters. 

The GALEX GR6\footnote{http://galex.stsci.edu/GR6} provides the cross-matched table (\texttt{xSDSSDR7}) against SDSS DR7. We search this table for the UV sources (detected both in FUV and NUV images) matched against the input optical catalog and find a total of 34,388 UV sources cross-matched to a total of 23,493 unique SDSS objectIDs. About half of the matched sources are observed in the GALEX All-sky Imaging Survey (AIS; $t_{exp} \approx 110~sec$), and the other half in the Medium Imaging Survey (MIS) of a single orbit exposure ($t_{exp} \approx 1500~sec$) or in the Deep Imaging Survey (DIS) of multiple orbits. Due to the short exposure time, the GALEX AIS data are highly biased toward star-forming galaxies, whereas the quiescent ETGs are well defined in deeper exposure images. The GALEX AIS data are inadequate for the demographic study of ETG populations at these redshifts and hence will be ignored hereafter.

For those ETGs detected in sufficiently deep exposure images with the GALEX, we analyze their emission line characteristics of the optical spectra. The emission line strengths in the SDSS DR7 spectra are measured by \citet{oh11}, and the sample galaxies are classified by the emission line ratios, [OIII]/H$\beta$ and [NII]/H$\alpha$ \citep{bal81,kew01,kau03}. \citet{oh11} provide the amplitude-to-noise ($A/N$) ratio at each emission line to set the analysis threshold. As our main purpose is to secure the $quiescent$ (i.e., no emission line) ETGs as stringently as possible, we loosen the threshold to $A/N=1$ and classify the sample galaxies into quiescent (33\%), star-forming (7\%), composite (12\%), and AGN (48\%) categories.

Fig.~1 shows the color--magnitude diagrams and color--color diagrams of the cross-matched ETG catalog. In the left panels, our sample of ETGs is expressed in the u, g, and r-band magnitudes and the colors sub-grouped according to the emission line characteristics. The separation among the different spectroscopic subgroups (especially between the quiescent and AGN-host galaxies) is much more effective in UV colors, as shown in the right panels. Most of the quiescent ($red$) galaxies are located in the ``UV red sequence'', whereas the blue envelop is occupied by star-forming ($blue$) galaxies. AGN-host ($green$) galaxies are spread out in-between in the UV color--color diagram. We adopt the modelMag (SDSS) and MAG\_AUTO (GALEX) as the total magnitudes of the galaxies in each band. Even though the source extraction apertures are not identically applied between the photometry of the optical and UV images, the aperture effect is small compared to the observational errors \citep[see][]{ree07}.

Some E+A galaxies \citep{got07} are found in our data and their UV colors are indicated by asterisks in Fig.~1. E+A galaxies are located over the wide color ranges ($0<FUV-NUV<3$, $2<FUV-r<8$), but clearly separate from the ``UV red sequence''. They appear to be on an evolutionary sequence in which the UV flux fades away after the starburst, and yet FUV declines much faster than NUV \citep[e.g.,][]{kav07a,cho09}.

\section{UV COLOR--COLOR RELATION OF QUIESCENT EARLY-TYPE GALAXIES}

\begin{figure}
\begin{center}
\epsscale{1.1}
{\plotone{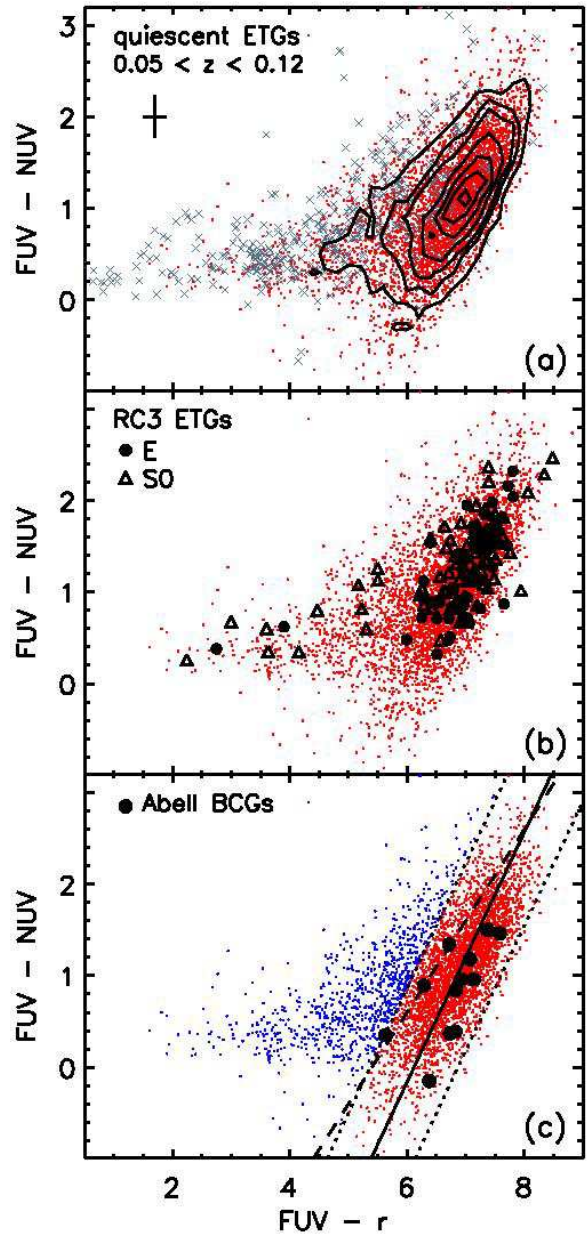}}
\end{center}
\caption{$FUV-NUV$ $vs.$ $FUV-r$ color--color diagram of the $quiescent$ ETG sample. $(a):$ Among 4025 spectroscopically quiescent ETGs, about 9\% of the sample (denoted by crosses) turn out to be early-type spiral galaxies, mergers, or morphologically disturbed galaxies in the visual inspection of SDSS images. The remaining 3677 morphologically cleaned ETGs consist of a well-defined ``UV red sequence'' (denoted by contours). $(b):$ The UV colors of the final sample of quiescent ETGs at $0.05<z<0.12$ are compared to the color distribution of the RC3 E (filled circles) and S0 (triangles) galaxies. $(c):$ The ``UV red seqence'' is empirically defined by the linear fit (solid line) with 1.5~mag color width in $FUV-r$ (dotted lines). The colors of Abell BCGs \citep[filled circles,][]{ree07} fall well into the sequence. The semi-empirical criterion of RSF galaxies ($NUV-r=5.4$) in \citet{sch07} is indicated by the dashed line for comparison.\label{fig2}}
\end{figure}

We examine the morphology of individual galaxies in the SDSS optical images for those 4025 $spectroscopically$ $quiescent$ ETGs. There are still some contaminants of early-type spiral galaxies, mergers, or morphologically disturbed galaxies in our sample. We find that about 9\% of our sample should be rejected in the final sample of quiescent ETGs. This number also includes some possible blends in the GALEX UV images in the case that there are close neighbors of blue dwarf galaxies around a target ETG. Fig.~2(a) shows the colors of such contaminants (denoted by crosses). Note that we eliminate them by the optical morphology only, being blind to their UV or optical colors. Most of those morphological contaminants enter the $quiescent$ sample because their optical spectra are taken within the small fiber aperture ($3\arcsec$ in diameter) toward the galactic centers, while they do contain young stars in the outer region judged from the strong extended UV emission. The remaining 3677 $morphologically$-$cleaned$ $quiescent$ ETGs consist of a well-defined ``UV red sequence'' as denoted by contours in Fig.~2(a), with the peak number density at $FUV-NUV \approx 1.1$ and $FUV-r \approx 7.0$. The UV colors of the final sample of quiescent ETGs at $0.05<z<0.12$ are very similar to the color distribution of the local ETGs (Fig.~2b). The GALEX RC3 E (filled circles) and S0 (triangles) galaxy data are from Donas et~al. (2007), and their V magnitudes are converted to r-band with $V-r=0.29$ assumed from the 12~Gyr elliptical galaxy model in \citet{ree07}. 

Based on the semi-empirical criterion of \citet{sch07}, 31\% of the quiescent ETGs in our sample are classified as the recent star formation (RSF) galaxies ($NUV-r<5.4$), which is consistent with the findings of \citet{kav07b} and \citet{sch07}. Note, however, that the observed UV colors of quiescent ETGs follow the sequence of a much steeper slope rather than the semi-empirical model line (i.e., a harder UV spectral shape at a given UVX amplitude). Using a sigma-clipping linear fit algorithm to the observed colors iteratively, we derive the color--color relation, $(FUV-NUV) = 1.36~(FUV-r) - 8.35$, for the quiescent ETGs, which indeed has a steeper slope than that of the semi-empirical model line, i.e., $NUV-r=5.4$. In Fig.~2(c), we empirically define the ``UV red sequence'' (red dots) from the linear fit to the observed colors (solid line) with 1.5~mag color width (which corresponds to 3 times the mean absolute deviation of the observed data from the fit line) in $FUV-r$ (dotted lines). The UV colors of the BCGs in Abell clusters (filled circles) fall well into this sequence. There are still a significant number (20\%, blue dots) of outliers from the UV red sequence with this empirical definition, and their additional UV flux should be attributed to the presence of small fractions ($\sim$1\% in mass) of young stars \citep{yi05}. The number fraction of such outliers is negligible ($<$ 5\%) in the brightest ($M_r < -22$) subgroup, but gradually increases toward the less bright galaxies ($\sim 30\%$ for $M_r \approx -21$ subgroup). We find no substantial difference between the lower redshift ($0.05<z<0.086$) and the higher redshift ($0.086<z<0.12$) subgroups.

All the colors of galaxies presented here are derived from the apparent total magnitudes in each filter with no other corrections applied, except for the Galactic extinction \citep{sch98,wyd07}. The effects of redshift ($0.05<z<0.12$) and evolution ($\Delta$t $\approx$ 1~Gyr) are small compared to the intrinsic color spreads among the sample galaxies and should have little impact on the general morphology of the color--color diagram. However, better statistics (e.g., the number fraction of RSF galaxies) require the sample galaxies to be at equivalent cosmological distances. More detailed demographic analysis is in preparation and will be presented in a future paper.

\section{MODEL COMPARISONS}

We employ the $\chi^2$ minimization technique of \citet{yi11} to find the best fit model spectra for the observed multiband photometry. Most of the observed data in Fig.~2 are well fitted by the two-component models in which the UV spectral shape is determined by the combination of the baseline old (12~Gyr) stellar spectra and the varying amount (0.01 -- 100\%) of a young (0.01 -- 10~Gyr) stellar component. However, the fitting accuracy is poor for the sample of quiescent galaxies in the lower right corner ($FUV-NUV<0.9$, $FUV-r>6.0$) of Fig.~2. The galaxy in this particular color domain has a very hard UV spectral shape (comparable to that of NGC~1399 which is one of the strongest UVX galaxies in the local universe), but its UVX amplitude is relatively small.

The observed UV colors are compared with other population synthesis models in Fig.~3. We first adopt the composite stellar population (CSP) model (t = 12~Gyr, $Z=Z_\odot$) of \citet{mar05} for the baseline model of bulk populations which does $not$ contain blue HB stars. Then, in order to simulate the UV color variations due to different types of UV sources, we add the simple stellar populations (SSPs) of $Z=0.1~Z_\odot$ for the young components (t = 0.1, 0.5, and 1.0~Gyr) and for the old component (t = 12~Gyr, 100\% blue HB) from \citet{con10} models. The recent models of \citet{chu11} are also considered for the helium-enhanced cases (Y = 0.33, 0.38, and 0.43). The flux weight of SSPs is assumed to be 10\% of the baseline CSP model in r-band. This weight translates roughly to 0.3\% (0.1~Gyr), 1.5\% (1.0~Gyr), and 10\% (12~Gyr) in stellar mass fraction. The top panel shows the UV--optical model spectral energy distributions (SEDs) for the various possible UV sources, and the UV colors of each toy model are derived from the CSP+SSP model SEDs redshifted to $z=0.086$ (median redshift of our sample) as shown in the bottom panel. 

\begin{figure}
\begin{center}
\epsscale{1.15}
\plotone{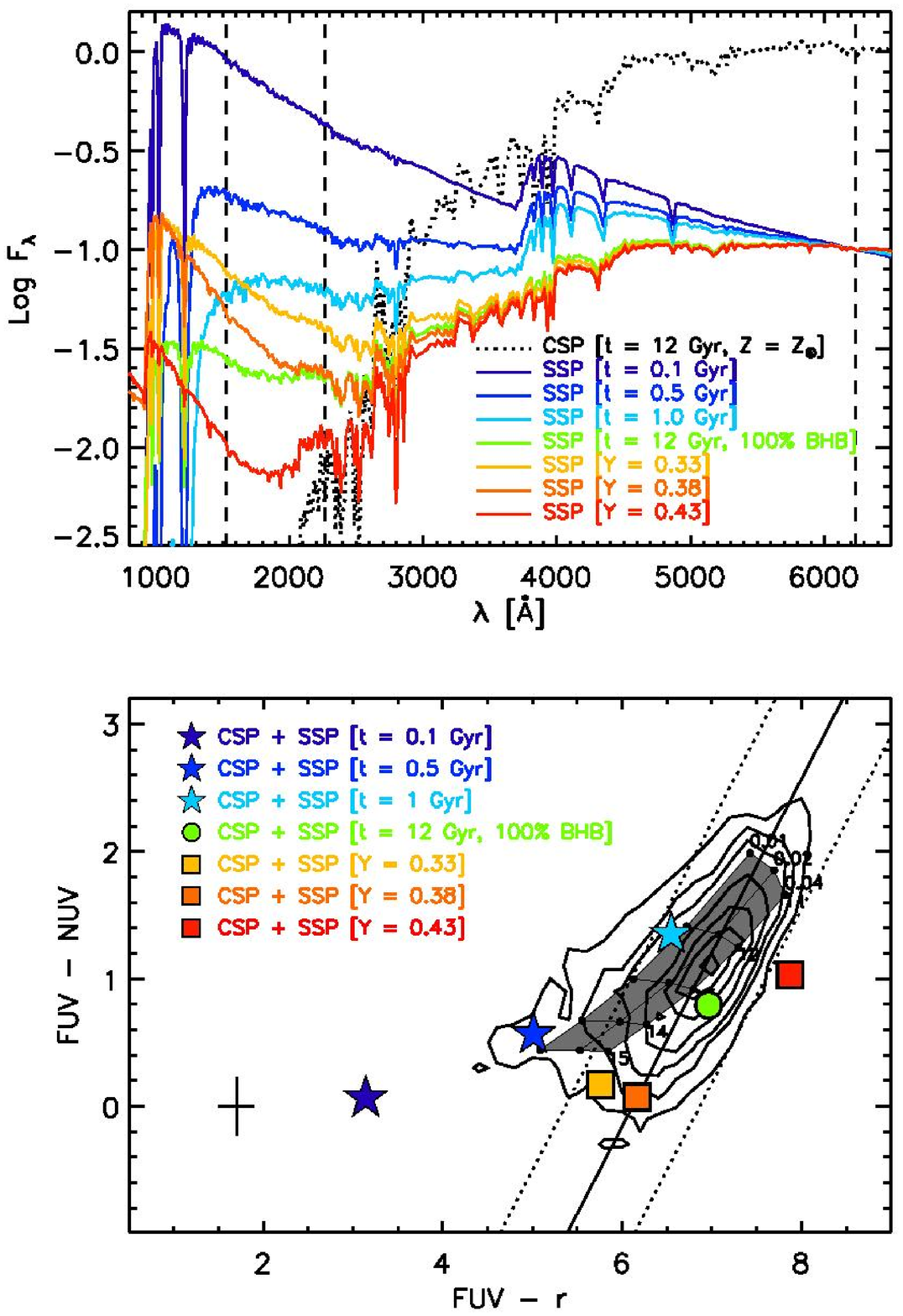}
\end{center}
\caption{$(Top) :$ UV--optical model SEDs. As the \citet{mar05} CSP model (12~Gyr, solar metallicity) does not contain blue HB stars, the SSP models ($Z=0.1~Z_\odot$) for young components (t = 0.1, 0.5, and 1.0~Gyr) and for old component (t = 12~Gyr, 100\% blue HB) from \citet{con10}, and for helium-enhanced cases (Y = 0.33, 0.38, and 0.43) from \citet{chu11} are additionally adopted to reproduce the UV spectra. The flux weight of SSP models is assumed to be 10\% of CSP in r-band. $(Bottom) :$ UV colors derived from the model SEDs redshifted to the median redshift of the sample. The combinations of CSP+SSP SEDs are as labeled. The canonical stellar population model grids in \citet{ree07} are denoted by the shaded region (t = 11 -- 15~Gyr, $<Z>$ = 0.01, 0.02, 0.04).
\label{fig3}}
\end{figure}

The shaded region in the bottom panel of Fig.~3 denotes the canonical models in \citet{ree07} where the mean temperature of HB stars smoothly varies according to metallicity and age \citep{lee94,yi99}. It is clear that the observed color spreads are much larger than the canonical model predictions spanning large ranges of age (11 -- 15~Gyr) and mean metallicity (0.5, 1.0, and 2.0 $Z_\odot$), and the difference cannot be attributed to observational errors or redshift effects. Moreover, the observed slope of the color--color relation is much steeper than the model predictions, implying that the UVX sources should have a harder UV spectral shape (i.e., hotter temperature) at a given amplitude systematically. An $ad~hoc$ assumption in the SED combination of \citet{mar05} CSP (no blue HB) and \citet{con10} SSP (100\% blue HB) models could produce a somewhat harder UV spectral shape at a given UVX amplitude (e.g., $FUV-r\approx7$) than the models in \citet{ree07}, and yet it is still hard to explain the UV spectral shape of the strongest UVX galaxies at $FUV-r\approx6$.

We test the recent model prediction of \citet{chu11} to determine whether the observed color spreads and the steep slope of the color--color relation might be reproduced with the helium-enhanced stellar populations. Since the helium-rich stars have lower masses at a given age, the mean temperature of helium-rich HB stars is much higher than the normal-helium populations. This effect is demonstrated well in Fig.~3. The combinations of the baseline CSP model and the helium-enhanced (Y = 0.33 and 0.38) SSP models would probably explain the UV spectra of the strongest UVX ($FUV-r\approx6$) galaxies under the $ad~hoc$ assumption that all the UV flux originates from the helium-enhanced HB stars only. An even higher helium abundance (e.g., Y = 0.43) seems unlikely to contribute in this color--color diagram as the HB stars of such high helium abundance are too hot and faint in the GALEX FUV bandpass. The UV spectral shapes of the helium-enhanced stars are clearly disentangled from those of the young stars which have a much stronger UV flux both in FUV and NUV at the same optical luminosity.

\section{CONCLUSION}

In summary, we made two main findings with the latest GALEX UV data of the ETGs in the nearby universe. 1) The ``old and dead'' ETGs consist of a well-defined sequence in UV colors in which the stronger UVX galaxies should have a harder UV spectral shape systematically. The UV color distributions of ETGs at $0.05<z<0.12$ are very similar to those observed for the local RC3 galaxies. The comparison with the population synthesis models indicates that the hot HB stars are responsible for the observed ``UV red sequence'', while the contamination of recently-formed young stars can be effectively discriminated in the UV color--color diagram. 2) The observed slope of the UV color-color relation of quiescent ETGs is too steep to be reproduced by the canonical stellar population models in which the UV flux is mainly controlled by age or metallicity parameters. Moreover, 2~mag of color spreads both in $FUV-NUV$ and $FUV-r$ appear to be ubiquitous among any subsets in distance or luminosity, which can hardly be explained by the canonical models with even large spreads in age ($>$ $\pm$2~Gyrs) and mean metallicity ($>$ $\pm$0.3~dex) assumed. This implies that the UVX in ETGs could be driven by yet another parameter which might be even more influential than age or metallicity. 

The recent debate on the helium-enhanced populations has an interesting implication for this issue. Observations suggest that there are some helium-enhanced subpopulations in the extended HB globular clusters \citep{dan04,nor04,lee05,pio05,rec06,lee07} and in the massive globular clusters in M87 \citep{soh06}. If present, such subpopulations would steepen the UV spectral slope systematically by increasing the mean temperature of helium burning stars, as demonstrated in Fig.~3. They would also enhance the UV color spreads among the ETG stellar populations with even small variations in the number fractions or in the amount of helium enhancement \citep[e.g.,][]{kav07c,chu11}. Whether the postulated helium enhancement in old stellar systems would occur also in galactic scales or in the intracluster medium of galaxy clusters \citep{pen09} is in question. Observational tests by \citet{lou11} and \citet{yi11} do not support the helium sedimentation hypothesis, while \citet{car11} argue that the helium abundance is a plausible candidate to explain the UV radial gradient. By the comparison between the UV color distribution and some toy models presented in this $Letter$, it is suggested that the observed steep slope of the UV color--color relation of ETGs would be better explained with the helium-enhanced subpopulations. The excessive spreads in the UV colors observed among the ETGs, besides the variations expected by the age/metallicity spreads, would be attributed to small variations in helium abundance or to the low-level star formation that is hard to be identified with the current survey depths and resolutions of the GALEX and SDSS data.

\acknowledgments{We thank Sukyoung Yi and Young-Wook Lee for their insights and discussions on the manuscript. We also thank the anonymous referee for a number of helpful comments and suggestions. This work was supported by the extragalactic research project, DREAM of KASI.}

\end{document}